\documentstyle[12pt,epsf]{article}

\begin{document}
\renewcommand{\thepage}{}

\begin{titlepage}


\title{
\hfill
\parbox{5cm}{
\normalsize UT-796\\
}\\
\vspace{2ex}\large
Large Lepton Mixing in a Coset-space Family Unification on
E$_7$/SU(5)$\times$U(1)$^3$
\vspace{2ex}}
\author{\large
J.~Sato\thanks{e-mail address:
 {\tt joe@hep-th.phys.s.u-tokyo.ac.jp}}\
and\
T.~Yanagida\thanks{e-mail address:
 {\tt yanagida@hep-th.phys.s.u-tokyo.ac.jp}}\\
 {\it Department of Physics,
   University of Tokyo,}\\
   {\it   Tokyo 113, Japan}}
\date{\today}

\maketitle

\begin{abstract}
We study a coset-space unification model for families based on\\
E$_7$/SU(5)$\times$U(1)$^3$.
We find that qualitative structure of quark and lepton mass matrices
in this model describes
very well the observation.
We stress, in particular, that the large mixing angle,
$\sin^22\theta_{\nu_\mu\nu_\tau} \simeq 1$, required for
the atmospheric neutrino oscillation reported by the SuperKamiokande collaboration,
is naturally obtained,
which is a consequence of unparallel family structure in the
present coset-space unification.\\
\end{abstract}



\end{titlepage}

\newpage
\renewcommand{\thepage}{\arabic{page}}


The problem of quark-lepton mass matrices is one of
the most important issues in particle physics. It is widely
expected that these mass matrices contain valuable
information on a more fundamental theory beyond the
standard model. Much experimental effort has been done
to determine the quark masses and mixing angles and as a
consequence most of the relevant angles and eigenvalues has been
obtained. On the other hand, very little is known for
neutrino masses and flavor mixing in the lepton sector.
There are only two experimental hints of the flavor mixing at present:
one is the well-known solar neutrino deficit\cite{Solar}
and the other the atmospheric neutrino anomaly\cite{Atm}.

A recent report on the atmospheric neutrino from the
SuperKamiokande collaboration\cite{SK}
has presented a convincing evidence that the atmospheric neutrino
anomaly is indeed due to neutrino oscillation. 
They have reported an asymmetry between up- and down-going 
$\nu_\mu$ fluxes. This up-down asymmetry indicates that
the $\nu_\mu$'s from overhead do not travel enough to oscillate,
while  the $\nu_\mu$'s coming from across the earth travel
sufficiently to oscillate. Such an asymmetry
has not be seen in the $\nu_e$ flux. Thus, this implies $\nu_\mu-
\nu_\tau$ oscillation with the mass difference 
$\delta m^2_{\nu_\mu\nu_\tau} \simeq 10^{-3}-10^{-2}$ eV$^2$\cite{SK}
which suggests $m_{\nu_\tau} \simeq (0.3-1)\times 10^{-1}$
eV provided $m_{\nu_\tau} > m_{\nu_\mu}$.

The above neutrino mass indicates the presence of 
right-handed neutrinos at the scale 
$10^{14}-10^{15}$ GeV\cite{seesaw}. The most natural 
theory accommodating such 
right-handed neutrinos is the SO(10) grand unified theory(GUT).
However, the SO(10) GUT seems to have a difficulty
to explain the other surprising fact, i.e. a very
large mixing angle $\sin^22\theta_{\nu_\mu\nu_\tau}\simeq 1$\cite{SK},
unless there is a huge hierarchy in Mayorana masses of 
right-handed neutrinos\cite{Rnu}. An alternative theory
accommodating naturally the right-handed neutrinos is the
coset-space family unification\cite{E71} in supersymmetric (SUSY)
GUT's. Coset-spaces based on E$_7$ are known as unique
choices to contain three families of quarks and leptons\cite{E72}.
Among them E$_7/$SU(5)$\times$U(1)$^3$
is the most interesting,
since it contains also three families of right-handed neutrinos
as Nambu-Goldstone (NG) multiplets\cite{YY}.

In this paper we point out that the observed large lepton mixing,
sin$^22\theta_{\nu_\mu\nu_\tau}\simeq 1$, is naturally
explained in the  E$_7/$SU(5)$\times$U(1)$^3$ model.

The  E$_7/$SU(5)$\times$U(1)$^3$ model\cite{E72,YY} contains
three families of ${\bf 10}_i + {\bf 5}^*_i + {\bf 1}_i$
($i=1-3$)
and one ${\bf 5}$ as NG multiplets. Here, the SU(5) is the usual
GUT gauge group.
Their quantum numbers under the unbroken subgroup are given in Table
\ref{charge}.

\begin{table}[ht]
\begin{center}
\begin{tabular}{c|ccc}
SU(5)
&\makebox[2cm]{U(1)$_1$}&\makebox[2cm]{U(1)$_2$}&\makebox[2cm]{U(1)$_3$}\\
\hline
{\bf 10}$_1$&0&0&4\\
{\bf 10}$_2$&0&3&-1\\
{\bf 10}$_3$&2&-1&-1\\
{\bf 5}$^*_1$&0&3&3\\
{\bf 5}$^*_2$&2&-1&3\\
{\bf 5}$^*_3$&2&2&-2\\
{\bf 1}$_1$&0&3&-5\\
{\bf 1}$_2$&2&-1&-5\\
{\bf 1}$_3$&2&-4&0\\
\hline
{\bf 5}&2&2&2
\end{tabular}
\caption{U(1) charges of the NG multiplets. The U(1)$_1$,
U(1)$_2$ and U(1)$_3$ are the unbroken U(1)'s of
coset-subspaces E$_7$/E$_6\times$U(1),
E$_6$/SO(10)$\times$U(1) and SO(10)/SU(5)$\times$U(1), respectively.}
\label{charge}
\end{center}
\end{table}

This model can not be quantized in the original form,
since there is a nonlinear-sigma model anomaly\cite{Anom,YY}.
However, this global obstruction is easily removed\cite{YY}
by introducing a matter multiplet ${\bf 5}^*$
which is also needed for an SU(5) gauge-anomaly
cancellation\cite{E72}. We assume that some explicit breaking induces an
invariant mass for the NG ${\bf 5}$ and this matter ${\bf 5}^*$
and we neglect them in our discussion.

In addition to the NG multiplets we introduce a pair of Higgs
multiplets ${\bf 5}_H$ and ${\bf 5}^*_H$. As long as
the global E$_7$ is exact these Higgs multiplets never
have Yukawa couplings to the NG quarks and leptons.
Thus, the observed hierarchy in quark-lepton mass
matrices is regarded as a consequence of a hierarchy 
in the explicit breaking of the global E$_7$.
This situation is very similar to that in the QCD,
where the mass hierarchy between NG pions and kaons
($m_K^2\gg m_\pi^2$) is originated from the hierarchy
in quark masses ($m_s\gg m_{u,d}$) which are explicit
breaking parameters of the chiral SU(3)$_L\times$SU(3)$_R$.

We consider three steps for the explicit breaking:
\begin{equation}
\begin{tabular}{ccccccc}
$ {\rm E}_7$ &$\longrightarrow $&$   {\rm E}_6$& $\longrightarrow$ &
${\rm SO(10)}$&$ \longrightarrow$ &${\rm SU(5)}$,\cr
&$\epsilon_0$&&$\epsilon_1$&&$\epsilon_2
$
\end{tabular}
\label{breaking}
\end{equation}
which leads to the mass hierarchy
\begin{eqnarray}
m_t \gg m_c \gg m_u \nonumber\\
m_b\gg m_s\gg m_d\\
m_\tau\gg m_\mu\gg m_e.\nonumber
\end{eqnarray}

To realize this hierarchy we assume that
the global E$_7$is broken explicitly by
the fundamental representation of E$_7, {\bf 56}$,
which contains six breaking parameters,
$\epsilon_0, \bar{\epsilon}_0,\epsilon_1, \bar{\epsilon}_1,
\epsilon_2, \bar{\epsilon}_2 $ that are all singlets of
SU(5). They carry U(1) charges as
\begin{eqnarray}
&\epsilon_0 (-3,0,0), &\bar{\epsilon}_0 (3,0,0)\nonumber\\
&\epsilon_1 (-1,-4,0), &\bar{\epsilon}_1 (1,4,0)\\
&\epsilon_2 (-1,-1,-5), &\bar{\epsilon}_2 (1,1,5)\nonumber
\end{eqnarray}
where the numbers in each parenthesis denote charges
of U(1)$_1\times$U(1)$_2\times$U(1)$_3$. The desired hierarchy
in eq.(\ref{breaking}) is represented by
\begin{equation}
\epsilon_0 \gg \epsilon_1 \gg
\epsilon_2.
\end{equation}

The structure of Yukawa couplings for the NG quarks and leptons
depends on U(1) charges of the Higgs {\bf 5}$_H$ and {\bf 5}$^*_H$.  
To determine them,
we consider that the Higgs multiplets  ${\bf 5}_H$ and
${\bf 5}_H^*$ belong to {\bf 27} of E$_6$ in
{\bf 133} of E$_7$. Then,
U(1) charges for the {\bf 5}$_H$ are given by
\begin{equation}
  {\bf 5}_H \ (2,2,2).
\end{equation}
The Higgs {\bf 5}$^*_H$ is a linear combination of two
{\bf 5}$^*$'s in {\bf 27} of E$_6$\footnote{
{\bf 27} of E$_6$ is decomposed to {\bf 16} + {\bf 10} +
{\bf 1} of SO(10). The {\bf 16} and {\bf 10} contain one {\bf 5} and
two {\bf 5}$^*$ of SU(5).} as
\begin{equation}
  {\bf 5}^*_H = \sin\theta {\bf 5}^*_{16} + \cos\theta {\bf 5}^*_{10}
\end{equation}
where U(1) charges for  {\bf 5}$^*_{16}$ and {\bf 5}$^*_{10}$
are given by\footnote{
The orthogonal combination of the {\bf 5}$_{16}^*$ and {\bf 5}$^*_{10}$
is assumed to have a GUT scale mass. We also assume that color
triplets in {\bf 5}$_H$ and {\bf 5}$^*_H$ receive a GUT scale mass
after the spontaneous breakdown of the SU(5) GUT.
This requires a fine tuning. We do not, however,
discuss this fine tuning problem here, since it is beyond the scope of this
paper.}
\begin{equation}
{\bf 5}^*_{16} (2,-1,3) \ \ {\rm and}
\ \  {\bf 5}^*_{10} (2,2,-2).
\end{equation}

We  now discuss Yukawa couplings for the quark
and lepton multiplets.
In general, Yukawa couplings are given
in a form $a_n\epsilon^n\psi\psi H$ where $\epsilon$, $\psi$
and $H$ stand for the explicit
breaking parameters, the NG multiplets and the Higgs
multiplets, respectively. By our choice of the U(1) charges for the explicit breaking
parameters and Higgs multiplets,
Yukawa couplings take the following form in the leading order of the
explicit breaking
parameters, $\epsilon$'s;
\begin{eqnarray}
  \label{Yukawas}
  W &=& W_U + W_D + W_E + W_\nu,\\
W_U &=& \sum_{ij} a_{ij} Y_{Uij} {\bf 10}_i {\bf 10}_j {\bf 5}_H,
\label{Wu}\\
W_D &=& W_E = \sum_{ij} b_{ij} Y_{D/Eij} {\bf 5}^*_i {\bf 10}_j {\bf 5}^*_H,
\label{Wde}\\
W_\nu &=& \sum_{ij} c_{ij} Y_{\nu ij} {\bf 5}^*_i {\bf 1}_j {\bf 5}_H,
\label{Wnu}
\end{eqnarray}
where $W_U,\ W_D,\ W_E$ and $W_\nu$ represent superpotentials
of Yukawa couplings for
up-type quarks, down-type quarks, charged leptons and neutrinos.
In these expressions $Y$'s are given by,\footnote{
One may wonder that in eq.(\ref{Yde}) the (3,1) element of $Y_{D/E}$,
has a term of $\epsilon_0\epsilon_1$.
We do not think that such a term
appears there, since in the limit $\epsilon_2 \rightarrow 0$, the global
SO(10) symmetry becomes exact and the {\bf 10}$_1$ is the true
NG multiplet which has no Yukawa interaction in the superpotential.
}$^,$
\footnote{Precisely speaking, our coset-space E$_7$/SU(5)$\times$U(1)$^3$
contains three dimensional parameters $f_0,\ f_1$ and $f_2$.
We assume  $f_0\sim f_1\sim f_2$ here, for simplicity. However,
even if it is not the case, one obtains the same form of
Yukawa couplings as in eqs.(\ref{Yu}), (\ref{Yde}) and (\ref{Ynu})
by redefining $\epsilon$'s as $\epsilon_i = \tilde\epsilon_i/f_i$
($i=$0,1,2) where $\tilde\epsilon_i$ are original dimensional parameters
for the explicit E$_7$ breakings.}
\begin{equation}
  \label{Yu}
  Y_U \simeq \left(
\begin{array}{ccc}
\epsilon_2^2&\epsilon_1\epsilon_2&\epsilon_0\epsilon_2\\
\epsilon_1\epsilon_2&\epsilon_1^2&\epsilon_0\epsilon_1\\
\epsilon_0\epsilon_2&\epsilon_0\epsilon_1&\epsilon_0^2\\
\end{array}
\right),
\end{equation}

\begin{equation}
  \label{Yde}
  Y_{D/E} \simeq \left(
\begin{array}{ccc}
\epsilon_1\epsilon_2\cos\theta&\epsilon_1^2\cos\theta
&\epsilon_0\epsilon_1\cos\theta\\
\epsilon_0\epsilon_2\cos\theta&\epsilon_0\epsilon_1\cos\theta
&\epsilon_0^2\cos\theta\\
\epsilon_0\epsilon_2\sin\theta&\epsilon_0\epsilon_1\sin\theta
&\epsilon_0^2\sin\theta\\
\end{array}
\right),
\end{equation}

\begin{equation}
  \label{Ynu}
  Y_\nu \simeq \left(
\begin{array}{ccc}
\epsilon_1^2&\epsilon_0\epsilon_1
&\epsilon_0\epsilon_2\\
\epsilon_0\epsilon_1&\epsilon_0^2&0\\
0&0&\epsilon_0^2
\end{array}
\right)
\end{equation}

We have assumed the E$_7$ representations for $\epsilon_i$, {\bf 5}$_H$
and {\bf 5}$^*_H$ to determine their U(1) charges. However, we
consider that this assumption is over statement since the E$_7$
is already spontaneously broken. What is relevant to our analysis is
only their charges of the unbroken subgroup SU(5)$\times$U(1)$^3$.
With this general consideration it is
impossible to estimate the coefficients 
$a_{ij},\ b_{ij}$ and $c_{ij}$  in eqs.(\ref{Wu}),
(\ref{Wde}) and (\ref{Wnu})
and hence we assume that they are of O(1) in this paper.

From the above Yukawa couplings in eqs.(\ref{Yu})
and (\ref{Yde}) we easily derive the following mass
relations;
\begin{eqnarray}
&\displaystyle
\frac{m_u}{m_c} \sim \frac{\epsilon_2^2}{\epsilon_1^2},\ 
&\frac{m_c}{m_t} \sim \frac{\epsilon_1^2}{\epsilon_0^2},\nonumber\\
  \label{massrelation}
&\displaystyle
\frac{m_e}{m_\mu} = \frac{m_d}{m_s} \sim \frac{\epsilon_2}{\epsilon_0}
\sin^{-1}\theta,
&\frac{m_\mu}{m_\tau} = 
\frac{m_s}{m_b} \sim \frac{\epsilon_1}{\epsilon_0}\sin\theta\cos\theta.
\end{eqnarray}
These relations describe very well the observed mass relations
provided that 
\begin{eqnarray}
  \label{eps}
\frac{\epsilon_1}{\epsilon_0} \sim 0.1,\ \ \frac{\epsilon_2}{\epsilon_0}
\sim 0.01 \ \ {\rm and}\ \ \tan\theta \sim 1.
\end{eqnarray}

We see that the Cabibbo-Kobayashi-Maskawa mixing angles for quarks
between the 1st and the 2nd,
the 2nd and the 3rd, and the 3rd and the 1st family are of the order
$\epsilon_2/\epsilon_1$,
$\epsilon_1/\epsilon_0$, and
$\epsilon_2/\epsilon_0$, respectively. It also describes the observed
mixing angles
 very well provided that the relations in eq.(\ref{eps}) are
satisfied.

We do not further mention details of the mass relations
since there should be corrections to the mass matrices in eqs.(\ref{Yu})
and (\ref{Yde}) from some higher
dimensional operators which
may affect masses for lighter particles significantly.
Otherwise, we have a SU(5) GUT relation, $m_d=m_e$,
which seems unrealistic\cite{PDG}.

So far, we have discussed the mass matrices for quarks and charged
leptons and found that the qualitative global structure of the obtained
matrices fits very well the observed mass spectrum
for quarks and charged leptons (except for $m_d=m_e$)
and mixing angles for quarks
if the relations in eq.(\ref{eps}) are satisfied\footnote{
The observed mass for the strange quark seems somewhat smaller
than the SU(5) GUT value\cite{PDG}.}.

We are now at the point to discuss neutrino masses and lepton mixings.
We assume that  Mayorana masses for right-handed
neutrinos $N_i$ are induced by SU(5) singlet Higgs
multiplets $\bar{s}_i({\bf 1})$.
We introduce two singlets $\bar{s_1}$({\bf 1}) and 
$\bar{s}_2$({\bf 1}) whose U(1) charges\footnote{
These $\bar{s_i}$({\bf 1}) are regarded as SU(5) singlet components
of {\bf 56} of E$_7$.} are given by
\begin{eqnarray}
  \label{singlets}
\bar{s}_1 (1,4,0)\ \ {\rm and}\ \ 
\bar{s}_2 (1,1,5).
\end{eqnarray}
Their vacuum expectation values, $\langle\bar{s_1}\rangle$ and
$\langle\bar{s_2}\rangle$ are expected to be of order
of the SU(5) GUT scale $\sim 10^{16}$ GeV.

Majorana masses for $N_i$ are induced from nonrenormalizable
interactions of a
form;\footnote{Other mass terms such as
$\epsilon^2 N_i N_j$ can be forbidden
by some chiral symmetry.}
\begin{equation}
W_N = \frac{\epsilon^2}{M_G} N_i N_j \bar{s}_k \bar{s}_l.
\end{equation}
Here, $M_G$ is the gravitational scale $M_G\simeq 2.4\times
10^{18}$ GeV.
Then, the matrix of the Majorana masses takes the following
form;\footnote{The mass term of the form $\epsilon^4 N_i N_j$
may produce a similar form to eq.(\ref{MnuR}) if $\bar{\epsilon}_0=0$
and $\bar{\epsilon}_1,\bar{\epsilon}_2 \ne 0$.}
\begin{equation}
  \label{MnuR}
  M_{\nu_R} =  \frac{1}{M_G}\left(
\begin{array}{ccc}
\epsilon_1^2 \bar{s_2}^2
&\epsilon_0\epsilon_1 \bar{s_2}^2&\epsilon_0\epsilon_1 \bar{s_1}
\bar{s_2}\\
\epsilon_0\epsilon_1  \bar{s_2}^2&
\epsilon_0^2 \bar{s_2}^2&\epsilon_0^2  \bar{s_1}\bar{s_2}\\
\epsilon_0\epsilon_1 \bar{s_1}
\bar{s_2}& \epsilon_0^2 \bar{s_1} \bar{s_2}&\epsilon_0^2 \bar{s_1}^2
\end{array}
\right),
\end{equation}
where all elements are multiplied by undetermined factors of O(1) 
like in the case for quarks and leptons.

The neutrino masses are given by\cite{seesaw}
\begin{equation}
  \label{numass}
  m_\nu \simeq m_{\nu_D} M_{\nu_R}^{-1} m_{\nu_D}^T,
\end{equation}
where
\begin{equation}
  \label{nudmass}
   (m_{\nu_D})_{ij} = c_{ij} Y_{\nu ij} \langle{\bf 5}_H \rangle.
\end{equation}
Three eigenvalues of the matrix in eq.(\ref{numass}) are of order,
$m_{\nu_1} \sim \epsilon^2_1 M_G \langle {\bf 5}_H \rangle^2/
\langle \bar{s}_2 \rangle^2,$
$m_{\nu_2} \sim \epsilon^2_0 M_G \langle {\bf 5}_H \rangle^2/
\langle \bar{s}_2 \rangle^2$ and $
m_{\nu_3} \sim \epsilon^2_0 M_G \langle {\bf 5}_H \rangle^2/
\langle \bar{s}_1 \rangle^2$.
It is remarkable that for $\langle {\bf 5}_H \rangle
\sim 100$GeV, $\epsilon_0 \sim 1$ and $\langle \bar{s}_i\rangle
\sim 10^{16}$GeV
we get the desired mass for neutrino $m_{\nu_i} \sim 0.1$ eV.

From the  Mikheev-Smirnov-Wolfenstein solution(MSW)\cite{MSW}
to the solar neutrino problem, we have\cite{BK,nurev}
\begin{equation}
  \delta m_{\nu_e\nu_\mu}^2 \simeq 10^{-6} - 10^{-5} {\rm eV}^2.
\end{equation}
We see that there are two choices
\begin{equation}
  \left(\frac{\langle \bar{s}_1 \rangle}{\langle \bar{s}_2 \rangle}
\right)^2
\sim 10^{-2}-10^{-1}\  {\rm or}\  
  \left(\frac{\langle \bar{s}_2 \rangle}{\langle \bar{s}_1 \rangle}
\right)^2
\sim 10^{-2}-10^{-1}
\end{equation}
to account for atmospheric and solar neutrino anomalies,
simultaneously. Thus, all off-diagonal elements of the 
diagonalization matrix for the neutrino mass matrix in eq.(\ref{numass})
are of O(0.1) in either cases.

However, it is very interesting that 
the mixing angle for lepton doublets
which mixes  charged leptons in the second and the third family
is of order $\tan\theta$ (see eq.(\ref{Yde})) and hence of the order 1.
This means, together with the above result,
that the weak mixing angle relevant for
$\nu_\mu$-$\nu_\tau$ oscillation can be so large,
$\sin^2 2\theta_{\nu_\mu\nu_\tau} \simeq 1$, as required for explaining
the observed atmospheric neutrino anomaly.
On the other hand, the mixing angle for $\nu_\mu-\nu_e$
oscillation is very small, $\theta_{\nu_\mu\nu_e} \sim $ O(0.1),
which may fit the small angle MSW solution\cite{BK,nurev} to the solar
neutrino problem.

In this paper we have shown that the coset-space family unification
on E$_7$/SU(5)$\times$U(1)$^3$ naturally accommodates the large
lepton mixing, $\sin^22\theta_{\nu_\mu\nu_\tau} \simeq 1$,
 necessary for explaining the atmospheric neutrino
anomaly reported by the SuperKamiokande collaboration\cite{SK}.
The main reason why we have a large mixing of the SU(2) lepton doublets
in the second and the third family is the twisted
structure of family. Namely, the {\bf 5}$^*$'s
in the second and the third family both live on the same coset-subspace
E$_7$/E$_6\times$U(1). On the other hand the {\bf 10}'s
in the third, the second and the first family live on the separate 
coset-subspaces, E$_7$/E$_6\times$U(1), E$_6$/SO(10)$\times$U(1)
and SO(10)/SU(5)$\times$U(1), respectively.
This unparallel family structure is an unique feature of
the present coset-space family unification.

\end{document}